\documentclass[11pt,fleqn]{article}

\usepackage{amsfonts,amssymb,amsmath}
\usepackage{epsf,epsfig,graphics,graphicx}

\usepackage{amsfonts,amssymb,cite}
\usepackage{epsf,graphicx}

\newcommand{\bls}[1]{\renewcommand{\baselinestretch}{#1}}

\def\noi{\noindent}

\newcommand{\Title}[1]{\noi {{\Large\bf #1}}\\[1ex]}

\def\Aunames#1{\noi{\bf #1}}
\def\auth#1{${}^{#1}$}
\def\Addresses#1{\medskip\noi \protect
	\begin{description}\itemsep -3pt {\it #1} \end{description}}
\def\addr#1#2{\item[${}^{#1}$]{\it #2}}

\newcommand{\Abstract}[1]{\vskip 2mm \begin{center}
        \parbox{16.4cm}{\small\noi #1} \end{center}\medskip}

\def\email#1#2{\footnotetext[#1]{e-mail: #2}\addtocounter{footnote}{1}}

\def\nq{\hspace*{-1em}}
\def\nqq{\hspace*{-2em}}
\def\nhq{\hspace*{-0.5em}}

\def\cm{\hspace*{1cm}}
\def\inch{\hspace*{1in}}

\def\Jl#1#2{#1 {\bf #2},\ }

\def\ApJ#1 {\Jl{Astroph. J.}{#1}}
\def\CQG#1 {\Jl{Class. Quantum Grav.}{#1}}
\def\DAN#1 {\Jl{Dokl. AN SSSR}{#1}}
\def\GC#1 {\Jl{Grav. Cosmol.}{#1}}
\def\GRG#1 {\Jl{Gen. Rel. Grav.}{#1}}
\def\JETF#1 {\Jl{Zh. Eksp. Teor. Fiz.}{#1}}
\def\JETP#1 {\Jl{Sov. Phys. JETP}{#1}}
\def\JHEP#1 {\Jl{JHEP}{#1}}
\def\JMP#1 {\Jl{J. Math. Phys.}{#1}}
\def\NPB#1 {\Jl{Nucl. Phys. B}{#1}}
\def\NP#1 {\Jl{Nucl. Phys.}{#1}}
\def\PLA#1 {\Jl{Phys. Lett. A}{#1}}
\def\PLB#1 {\Jl{Phys. Lett. B}{#1}}
\def\PRD#1 {\Jl{Phys. Rev. D}{#1}}
\def\PRL#1 {\Jl{Phys. Rev. Lett.}{#1}}

\def\al{&\nhq}
\def\lal{&&\nqq {}}
\def\eq{Eq.\,}
\def\eqs{Eqs.\,}
\def\beq{\begin{equation}}
\def\eeq{\end{equation}}
\def\bear{\begin{eqnarray}}
\def\bearr{\begin{eqnarray} \lal}
\def\ear{\end{eqnarray}}
\def\earn{\nonumber \end{eqnarray}}

\def\nnn{\nonumber\\ \lal }

\def\yy{\\[5pt] {}}
\def\yyy{\\[5pt] \lal }
\def\eql{\al =\al}

\def\sequ#1{\setcounter{equation}{#1}}

\def\dst{\displaystyle}
\def\tst{\textstyle}
\def\fracd#1#2{{\dst\frac{#1}{#2}}}
\def\fract#1#2{{\tst\frac{#1}{#2}}}
\def\Half{{\fracd{1}{2}}}
\def\half{{\fract{1}{2}}}

\def\e{{\,\rm e}}
\def\d{\partial}

\def\im{\mathop{\rm Im}\nolimits}

\def\diag{\mathop{\rm diag}\nolimits}

\def\const{{\rm const}}
\def\eps{\varepsilon}
\def\ep{\epsilon}

\def\mn{_{\mu\nu}}
\def\MN{^{\mu\nu}}
\def\mN{_\mu^\nu}

\def\wh{wormhole}
\def\whs{wormholes}
\def\bh{black hole}
\def\bhs{black holes}
\def\bu{black universe}
\def\bus{black universes}
\def\sph{spherically symmetric}
\def\ssph{static, spherically symmetric}
\def\asflat{asymptotically flat}

\def\GR{general relativity}
\def\KS{Kantowski-Sachs}

\def\Schr{Schr\"odinger}

\bls{1.02}
\begin{document}
\twocolumn[

\Title{Magnetic wormholes and black universes with invisible ghosts}

\Aunames {K. A. Bronnikov\auth{a,b,c,1} and P. A. Korolyov\auth{b,2} }

\Addresses{
\addr a {Center of Gravitation and Fundamental Metrology,
         VNIIMS, Ozyornaya St. 46, Moscow 119361, Russia}
\addr b {Institute of Gravitation and Cosmology,
         PFUR, Miklukho-Maklaya St. 6, Moscow 117198, Russia}
\addr c {I. Kant Baltic Federal University, Al. Nevsky St. 14,
         	Kaliningrad 236041, Russia}
      }

\Abstract
{We construct explicit examples of globally regular static, spherically
 symmetric solutions in general relativity with scalar and electromagnetic
 fields describing traversable wormholes with flat and AdS asymptotics
 and regular black holes, in particular, black universes. (A black universe
 is a regular black hole with an expanding, asymptotically isotropic
 space-time beyond the horizon.) The existence of such objects requires
 invoking scalars with negative kinetic energy (``phantoms'', or ``ghosts''),
 which are not observed under usual physical conditions. To account for that,
 the se-called ``trapped ghosts'' were previously introduced, i.e., scalars
 whose kinetic energy is only negative in a restricted strong-field region
 of space-time and positive outside it. This approach leads to certain
 problems, including instability (as is illustrated here by derivation
 of an effective potential for spherical pertubations of such systems).
 In this paper, we use for model construction what we call ``invisible
 ghosts'', i.e., phantom scalar fields sufficiently rapidly decaying in the
 weak-field region. The resulting configurations contain different numbers
 of Killing horizons, from zero to four.
 }

] 
\email 1 {kb20@yandex.ru}
\email 2 {korolyov.pavel@gmail.com}

\section{Introduction}

  The existence of the so-called exotic matter, violating the weak and null
  energy conditions, is favored by modern cosmological observations allowing
  for the ratio of pressure to energy density, $w = p/\rho < -1$ (see, e.g.,
  \cite{planck15} and references therein).
  This is one of the reasons for the recent interest
  in the construction and properties of wormhole configurations in general
  relativity and its extensions (see, e.g.,
  \cite{vis-book, lobo-rev, we-book, we-14} for reviews),
  since, as is well known, it is the necessity of exotic matter that makes a
  fundamental problem in \wh\ construction \cite{hoh-vis}.

  It has also been discovered that if one admits the existence of exotic
  matter, for example, in the form of phantom scalar fields, then, in
  addition to \whs, there can appear quite a number of other interesting and
  unusual configurations, such as different types of regular black holes
  and, among them, the so-called \bus. The latter look in the static region
  basically the same as ``ordinary'' \bhs\ in \GR, but beyond the horizon,
  instead of a singularity, they contain an expanding universe which
  ultimately becomes isotropic and can be asymptotically de Sitter at large
  times \cite{bu1, bu2}.

  Since no exotic matter or phantom fields have been detected under usual
  physical conditions, it is desirable to avoid the emergence of such fields
  in an asymptotic weak-field region. To that end, it has been suggested
  \cite{trap1, trap2, trap3} to use a special kind of fields, named
  ``trapped ghosts'', which have phantom properties only in some restricted
  strong-field region and satisfy the standard energy conditions in the
  remaining part of space. With such a field, a variety of solutions have
  been obtained, including regular black holes, black universes and
  traversable wormholes.

  In all these models, the kinetic energy density smoothly passes zero at
  some scalar field value $\phi= \phi_0$, being negative at $\phi> \phi_0$.
  This transition point has certain undesirable properties, in particular,
  if we consider perturbations of a \ssph\ configuration with such a field,
  the corresponding effective potential has a singularity which should in
  general lead to a violent instability.

  Trying to avoid these problems, in this paper we consider wormhole and
  black universe models without a trapped ghost but use, instead,
  a superposition of two scalar fields, a phantom one and a canonical one,
  requiring a sufficiently rapid decay of the phantom field in the
  weak-field region. We call this design an ``invisible ghost''.

  We use the electromagnetic field as one more source of gravity. As in
  \cite{trap3}, we deal with static, spherically symmetric space-times,
  therefore the only kinds of electromagnetic fields are a radial electric
  (Coulomb) field and a radial magnetic (monopole) field. For the latter,
  it is unnecessary to assume the existence of magnetic charges
  (monopoles): in both wormholes and black universes a monopole magnetic
  field can exist without sources due the space-time geometry. In the
  wormhole case it perfectly conforms to Wheeler's idea of a ``charge
  without charge'' \cite{wheeler}, and this charge can be both electric and
  magnetic.

  One of the motivations for including the electromagnetic field into
  consideration is that by modern observations there can exist a global
  magnetic field up to $10^{-15}$ Gauss, causing correlated orientations of
  sources remote from each other \cite{o-mag}, and some authors admit a
  possible primordial nature of such a magnetic field.

  The results obtained here show that a superposition of phantom
  and canonical scalar fields combined with an electromagnetic field can
  support wormholes, black universes and other kinds of regular black holes
  without a center. Actually, the set of possible types of geometry
  coincides with that obtained previously with the aid of a pure phantom
  scalar \cite{pha-mag} or a trapped ghost \cite{trap3}.

  The paper is organized as follows. In Section 2 we present the basic
  equations and make some general observations. In Section 3 we obtain
  examples of wormhole and regular \bh\ configurations supported by a
  superposition of two scalar fields, a phantom one and a canonical one.
  The Appendix presents the form of the effective potential for radial
  perturbations of \ssph\ field systems, more general than considered here,
  including those with a trapped ghost.

\section{Basic equations}

  Consider a field system with the action
\bearr                                  			\label{1}
	{\cal S} = \frac{1}{16\pi}\int \sqrt{-g} d^4 x
	\Big[ R + 2h_{ab}( \d\phi^a,\d\phi^b)
\nnn \inch
	- 2V(\phi^a) - F\mn F\MN \Big],
\ear
  in a \ssph\ space-time, where
  $R$ is the scalar curvature, $g = \det(g\mn)$,
  $F\mn$ is the electromagnetic field tensor,
  $\{\phi^a\}$ is a sigma-model type set of scalar fields,
  $h_{ab} = h_{ab}(\phi^a)$ is a nondegenerate target space metric,
  $(\d\phi^a,\d\phi^b) \equiv g^{ \mu \nu} \d_\mu\phi^a \d_\nu\phi^b$,
  and $V(\phi^a)$ is an interaction potential.

  The metric can be written as
\beq                                   \label{2}
       ds^2 = A(u)dt^2 - \frac{du^2}{A(u)} - r^2(u)d \Omega^2,
\eeq
  where we use the so-called quasiglobal gauge $g_{00}g_{11}=-1$, and
  $d \Omega^2=(d \theta^2+\sin^2 \theta d \varphi^2)$ is the linear element
  on a unit sphere.

  Our interest is in \wh\ and \bu\ solutions, describing nonsingular
  configurations without a center. Hence we assume that the range of $u$
  is $u \in \mathbb{R}$, where both $A(u)$ and $r(u)$ are regular, $r > 0$
  everywhere, and $r \to \infty$ at both ends. We also require $r(u)
  \approx |u|$ as $u \to \pm \infty$, which is in agreement with possible
  flat, de Sitter or AdS symptotic behaviors at large $r$.

  The existence of two asymptotic regions with $r \sim |u|$ implies that
  there is at least one regular minimum of $r(u)$ at some $u=u_0$,
  at which (the prime stands for $d/du$)
\beq                                                \label{3}
	r = r_0 > 0, \quad\ r'=0, \quad\  r''>0.
\eeq
  The necessity of violating the weak and null energy conditions (WEC and
  NEC) at such minima follows from the Einstein equations. Indeed, one of
  them reads (see \eq (\ref{11}) below)
\beq                                                     \label{4}
	2 A r''/r = -(T^t_t - T^u_u),
\eeq
  where $T\mN$ are components of the stress-energy tensor (SET).
  If a minimum of $r$ occurs in an R-region (i.e., $A > 0$), it is a throat.
  The condition $r''> 0$ implies, according to (\ref{4}), $T^t_t-T^u_u < 0$,
  or, in conventional terms, ($T^t_t = \rho$ and $-T^u_u = p_r$, the energy
  density and radial pressure, respectively) $\rho + p_r < 0$, which
  manifests NEC violation. It is a simple proof for static, spherically
  symmetric wormhole throats (\cite{thorne}; see also \cite{we-book}).

  If a minimum of $r$ occurs in a T-region ($A < 0$), it is not a throat but
  a bounce in the time evolution of one of the scale factors in a \KS\
  cosmology (the other scale factor is $[-A(u)]^{1/2}$). In a T-region $t$
  is a spatial coordinate, so $-T^t_t = p_t$ is the corresponding pressure,
  while $T^u_u = \rho$; however, the condition $r''> 0$ in (\ref{4}) leads
  to $\rho + p_t < 0$, again violating the NEC. In the intermediate case of
  $A = 0$ (a horizon) at a minimum of $r$, the condition $r''> 0$ should
  also hold in its vicinity, with all its consequences. Thus a minimum of
  $r$ always implies a NEC (and hence WEC) violation.

  The Einstein field equations can be written as
\beq                                              \label{7}
	R\mn = 2h_{ab} \d\mu\phi^a \d_\nu\phi^b - g\mn V + T\mn [F];
\eeq
  where $T\mn[F]$ is the SET of the electromagnetic field.
  Nonzero components of $F\mn$ compatible with the metric (\ref{2}) are
  $F_{01} = -F_{10}$ (a radial electric field) and $F_{23} = -F_{32}$ (a
  radial magnetic field), and the Maxwell equations lead to
\beq                                                      \label{8}
      F_{01}F^{01}=-q_{e}^2 /r^4(u), \quad\  F_{23}F^{23}=q_m^2/r^4 (u),
\eeq
  where the constants $q_{e}$ and $q_{m}$ are the electric and magnetic
  charges, respectively. The corresponding SET is
  ($q := \sqrt{q^2_e + q^2_m}$)
\beq  	                                           \label{9}
	T\mN [F] = \frac{q^2}{r^4 (u)} \diag(1,1,-1,1).
\eeq

  As to the scalar fields, let us assume that there are two of them, a
  usual, canonical field $\phi(u)$, and a phantom one, $\psi(u)$, able to
  provide NEC and WEC violation, so that $h_{ab} = \diag (1, -1)$. For the
  potential we make the simplest assumption
\beq                                \label{6}
	V = V_\phi(\phi) + V_\psi(\psi),
\eeq
  thus both fields are self-interacting but do not directly interact with
  each other. Then the set of equations to be solved can be written as
\bearr                    		\label{10}
	(A'r^2)' = -2r^2 V + 2 q^2/r^2,
\yyy					\label{11}
	r''/r = - \phi'^2  + \psi'^2,
\yyy					\label{12}
	A(r^2)''-r^2 A'' = 2- 4q^2/r^2,
\yyy                            	\label{s1}
	2\square \phi = dV_\phi/d\phi,
\yyy                            	\label{s2}
	2\square \psi = -dV_\psi/d\psi,
\ear
  where $\square$ is the d'Alembert operator: for any $f(u)$,
  $\square f = -r^{-2} [A r^2 f']'$. The unknowns are $\phi(u)$, $\psi(u)$,
  $A(u)$, $r(u)$, as well as the potentials $V_{\phi}(\phi)$ and
  $V_{\psi}(\psi)$. \eq (\ref{12}) can be integrated giving
\beq     \nhq                            \label{14}
      r^4 B'(u) = -2u + 4q^2  \int \frac{du}{r^2 (u)},
	\quad  B(u) \equiv \frac{A}{r^2}.
\eeq

\section{Examples of models with an invisible ghost}

  It is hard to solve \eqs (\ref{10})--(\ref{s2}) with given potentials
  $V_\phi$ and $V_\psi$. Instead, following the lines of \cite{bu1, trap1,
  trap2, trap3, pha-mag}, we try to find examples of interest using the
  inverse problem method: specifying the functions $r(u)$ and $\phi(u)$, we
  find all other unknowns from the field equations. Given the function
  $r(u)$ and the charge $q$, the redshift function $A(u)$ is found from
  (\ref{12}), and the summed potential $V(u)$ from (\ref{10}). The phantom
  field $\psi(u)$ can be found from (\ref{11}) provided $\phi(u)$ is known.
  Lastly, the separate potentials $V_\phi$ and $V_\psi$ are found using
  (\ref{s1}) and (\ref{s2}).

  We can choose the function $r(u)$, providing the opportunity of \wh\ and
  \bu\ configurations, in the same form as in \cite{trap3}:
\beq           				\label{15}
	r(u)=a \frac{x^2 +1}{\sqrt{x^2 +n}},  \qquad  n = \const > 2,
\eeq
  where $x=u/a$, and $a > 0$ is an arbitrary constant with the dimension of
  length. Hence,
\beq                \label{16}
	r''(u) = \frac{1}{a} \frac{x^2 (2-n) + n(2n-1)}{(x^2 +n)^{5/2}},
\eeq
  so that $r'' > 0$ at $x^2  < n(2n-1)/(n-2)$ (a behavior compatible
  with a phantom scalar) and $r'' < 0$ at larger $|x|$. It is also clear
  that $r \approx a |x|$ at large $|x|$. This ensures a negative kinetic
  energy density (see the r.h.s. of (\ref{11})) at low values of $x$ and
  a positive one in the rest of space. We will take $a=1$, which is in
  essence a choice of a length unit. The values of $r$, $q$, $m$ ($m$ is the
  Schwarzschild mass in our geometrized units), having the dimension of
  length, thus become dimensionless but are actually expressed in terms of
  $a$. The quantities $B(u)$ and $V(u)$ and others having the dimension
  (length)$^{-2}$, are expressed in units of $a^{-2}$. The quantities
  $A(u)$, $\phi(u)$, $\psi(u)$ are dimensionless.

  From \eq (\ref{14}) it follows
\bearr                            		\label{17}
    B'(x) =  \frac{(x^2 + n)^2}{(x^2+1)^4}
    \biggl[ 6p - 2x
\nnn\ \ \
    + 2 q^2
       \biggl(\frac{(n-1) x}{1 + x^2} + (n+1) \arctan x\biggr)\biggr],
\ear
\begin{figure*}
\centering
\includegraphics[width=55mm]{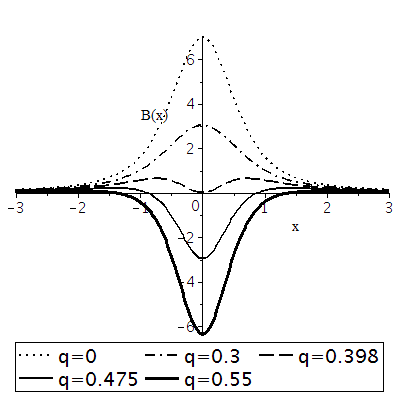}
\ \
\includegraphics[width=55mm]{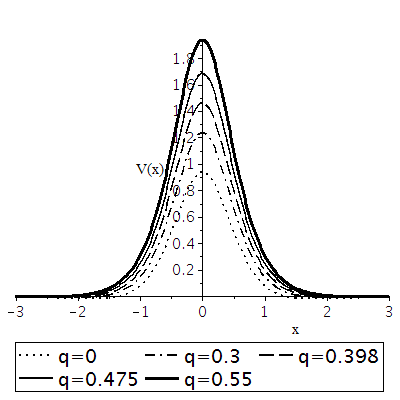}
\ \
\includegraphics[width=55mm]{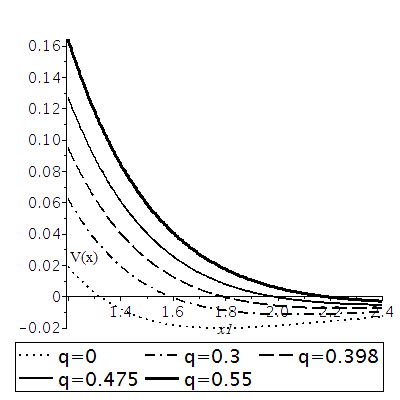}\\
	 a \hspace{55mm} b \hspace{55mm} c \hfill\\
\medskip
\caption{\small
  	The function $B(x)$ (a) and the potential $V(x)$ (b, c) for
  	symmetric configurations with $0 \leq q < 1$. Panel c shows that
	there is necessarily a region with negative values of $V$.
	}
\end{figure*}
  where $p$ is an integration constant. Further integration can be performed
  analytically but the resulting expressions for $B(x)$ and other quantities
  are too cumbersome. For definiteness, we further choose $n = 4$ (one can
  check that other values $n > 3$ do not qualitatively change the solution),
  then
\bearr                              \label{18}
	 B(u)= - \frac{27}{4} \frac{q^2 }{(x^2 +1)^4 }
\nnn
	 + \frac{1}{2} \frac{30 q^2 x \arctan(x)- 7 q^2 9 mx
	 	-6 \pi qx + 6}{(x^2 +1)^3}
\nnn \nq
	 + \frac{3}{16} \frac{180 q^2 x \arctan x +37 q^2 + 54 mx
	 	-36\pi qx + 16}{(x^2 +1)^2 }
\nnn
	 + \frac{1}{16} \frac{890q^2 x\arctan x + 445 q^2+267 mx}{x^2 +1}
\nnn
     	 + \frac{1}{16}\frac{ - 178\pi qx + 16}{x^2 +1}
     		+ \frac{445}{16} q^2  \arctan^2 x
\nnn
	 + \frac{89}{16} \arctan x (-3 m+2 \pi q) - B_0,
\ear
  with an integration constant $B_0$. Assuming asymptotic flatness as
  $x \to + \infty$, that is, $A \to 1$, we should require $B = A/r^2 \to 0$,
  which leads to
\beq                                                           \label{19}
	B_0 = \frac{445}{64}\pi^2 q^2 -\frac{89}{16} \pi^2 q
			+ \frac{267}{32} \pi m
\eeq
  Comparing the asymptotic expression $A(x) = 1 - 2m / x + o(1/x)$ with
  $A(x) = Br^2$ following from (\ref{17}), the parameter $p$ is related to
  the Schwarzschild mass $m$ and the charge $q$:
\beq                                                           \label{20}
	p = m - \frac{2}{3} \pi q^2
\eeq
  Thus $B$ is a function of $x$ and two parameters, the mass $m$ and the
  charge $q$. Furthermore, using (\ref{10}), we find the potential $V(x)$
  depending on the parameters $m$ and $q$.
\begin{figure*}
\centering
\includegraphics[width=57mm]{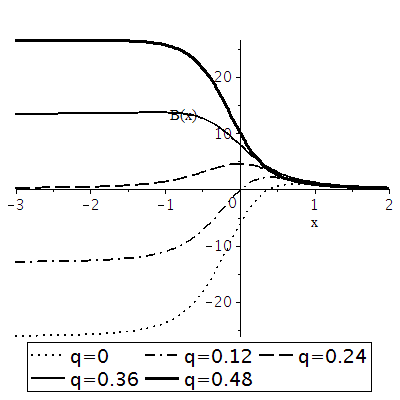}
\ \
\includegraphics[width=57mm]{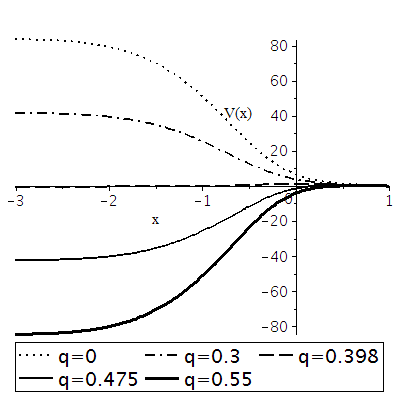}
\ \
\includegraphics[width=57mm]{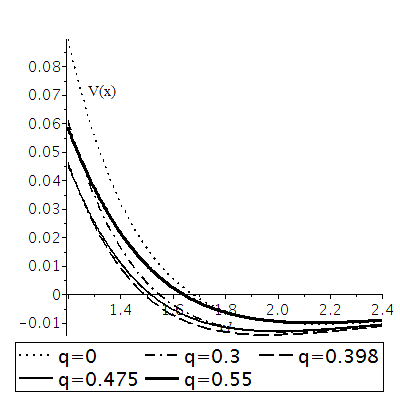}\\
	 a \hspace{55mm} b \hspace{55mm} c \hfill\\
\medskip
\caption{\small
  	The function $B(x)$ (a) and the potential $V(x)$ (b, c) for
 	some asymmetric configurations with $0 \leq q < 1$. Panel c
	shows the behavior of $V$ at comparatively large $x$ invisible
	in panel b.}
\end{figure*}

  At the other extreme, $x \to -\infty$, possible values $B(-\infty) < 0$
  correspond to de Sitter asymptotic behavior (dS), and, since it is a
  T-region, it is an expanding or contracting cosmology, in other words, we
  obtain a \bu.

  If $B(- \infty) = 0$, which happens if $p=0$, the metric is \asflat, and
  the whole solution is symmetric with respect to the surface $x=0$ since
  both $B(x)$ and $r(x)$, as well as the potential $V(x)$ are even functions.
  In this case, the mass and charge are related by $m = (2/3) \pi q^2$.

  Lastly, we obtain an anti-de Sitter (AdS) asymptotic if $B(- \infty) > 0$.
  The resulting solutions can be classified according to the asymptotic
  behaviors at the two infinities and the number and kind of horizons
  that correspond to zeros of $B(x)$.

\begin{figure*}
\centering
\includegraphics[width=60mm]{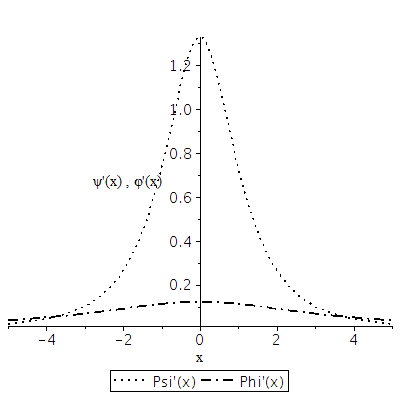}
\ \ \
\includegraphics[width=60mm]{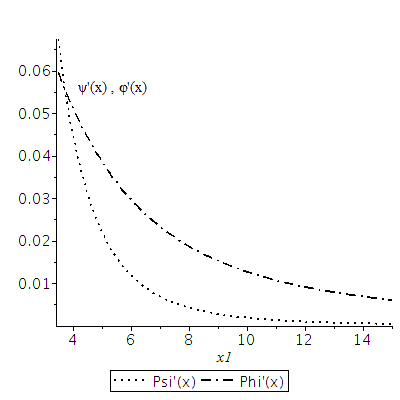}
\hfill
\caption{\small
	The quantities $\phi'$ and $\psi'$ in the strong and weak
	field regions}
\end{figure*}

\begin{figure*}
\centering
\includegraphics[width=60mm]{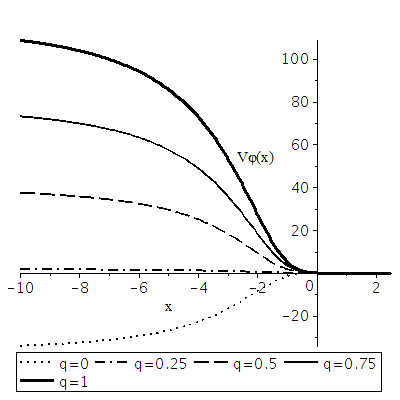}
\ \ \
\includegraphics[width=60mm]{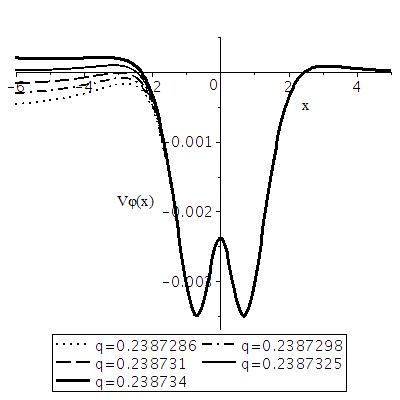}
\hfill\\
\caption{\small
	The potential $V_\phi(x)$ for $m=1/2$ and some values of the charge
	$q$.  The right panel demonstrates the detailed behavior of
	$V_\phi (x)$ close to the critical value $q = q_0 = 0.238673$, at
	which the number of its zeros changes. There is a very narrow range
	of $q$, $q_0 < q < 0.238733$, where $V_\phi (x)$ has two zeros
	at negative values of $x$.
	}
\end{figure*}

\begin{figure*}
\centering
\includegraphics[width=60mm]{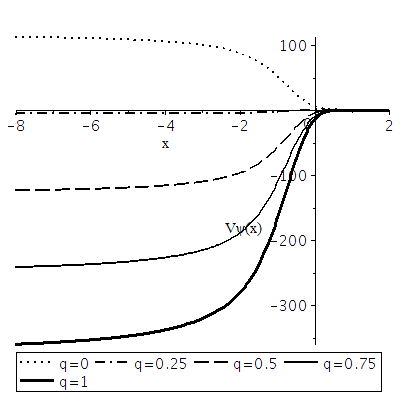}
\ \ \
\includegraphics[width=60mm]{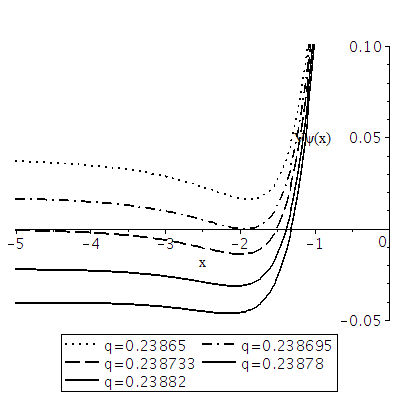}
\caption{\small
	The potential $V_\psi$ for $m=1/2$ and the same values of $q$ as in
	Fig.\,4. The right panel shows the behavior of $V_\psi (x)$ at
	$q$ close to $q_0$.}
\end{figure*}

  The possible kinds of geometries obtained have the same qualitative
  features as those discussed in \cite{pha-mag, trap3}, so we will not
  describe them here in full detail. In particular, we refer to \cite{pha-mag}
  for a description of the corresponding global causal structures and
  Carter-Penrose diagrams, which (since $r(x)\approx |x|$ at both
  infinities) are completely determined by the zeros of $B(x)$ and the signs
  of its asymptotic values.

  In the case of symmetric, twice asymptotically flat solutions we have the
  following types of geometries (see Fig.\,1): (i) a \wh\ ($0 \leq |q|
  \lesssim 0.398$), (ii) an extremal regular black hole with a single horizon
  $(q \approx 0.398)$, and (iii) a non-extremal regular black hole with two
  simple horizons $(q \gtrsim 0.398)$ (Fig.\,1). It is of interest that
  small changes of $q$ near the critical value $\approx 0.398$ drastically
  change $B(x)$ whereas the function $V(x)$ changes very little. One can
  also note that at large $|x|$ the potential $V(x)$ has small negative
  values (Fig.\,1c) while it is comparatively large and positive at small
  $x$.

  Asymmetric configurations, \asflat\ at $x=\infty$, are slightly more
  diverse. Fig.\,2 shows some of them under the assumption $m=1$ adopted for
  definiteness. Note that in both figures 1a and 2a all curves $B(x)$
  approach $x = \infty$ from positive since there is an R-region and
  Schwarzschild asymptotic.

  For the scalar field $\phi$, by analogy with \cite{trap3}, we assume
  the form
\beq                                                  	   \label{22}
	\phi(x) = K\arctan (Lx),
\eeq
  where $K$ and $L$ are adjustable constants. By (\ref{22}) and (\ref{11}),
\bearr                                                     \label{23}
	\phi'^2 (x) = \frac{K^2 L^2}{(L^2 x^2 +1)^2 },
\yyy 							\label{25}
 	\psi'^2 (x) = \frac{28- 2 x^2}{(x^2 +1)(x^2 +4)^2 }
		+ \frac{K^2 L^2 }{(L^2 x^2 +1)^2 }.
\ear

  Let us choose $K$ and $L$ in such a way as to make the phantom field
  $\psi$ decay at large $x$ more rapidly than $\phi$. From (\ref{25})
  we have
\beq  \nq
       \psi'^2 (x) = \frac{K^2- 2L^2}{L^2 x^4}
       + \frac{46L^4 -2K^2}{L^4 x^6} + O(x^{-8}),
\eeq
  hence taking
\beq                                                    \label{KL}
  	K = 2/\sqrt{23}, \quad\ L = \sqrt{2/23},
\eeq
  we obtain $\psi'\sim x^{-4}$ whereas $\phi' \sim x^{-2}$ at large $|x|$,
  as required. We use the values (\ref{KL}) in our numerical calculations.
  The behavior of $\phi'(x)$ and $\psi'(x)$ is shown in Fig.\,3.

    The field potential $V_{\phi}(x)$ is found from \eq (\ref{s1})
  under the boundary condition $V_\phi (\infty) =0$. The other potential
  $V_\psi$ is obtained as
\beq                      \label{32}
	V_\psi = V - V_\phi,
\eeq
  or it can be equivalently found from \eq (\ref{s2}). The expressions for
  the potentials are found analytically but are too bulky to be presented
  here, so we only show their plots for some values (Figs.\,4 and 5),
  again putting $m=1$. Other values of $m$ do not add any new qualitative
  features of the solution.

  Summarizing, we have obtained examples of \wh, regular \bh\ and \bu\
  solutions to the field equations with two scalars $\phi$ (canonical)
  and $\psi$ (phantom), in which the phantom field comparatively quickly
  decays at infinity.

\section*{Appendix: linear perturbations}
\sequ {0}
\def\theequation{A\arabic{equation}}
\def\da{\delta\alpha}
\def\db{\delta\beta}
\def\df{\delta\phi}
\def\dg{\delta\gamma}
\def\Veff{V_{\rm eff}}

  Consider an action with a single scalar field, but more general than
  (\ref{1}) in that we admit a dilatonic-type interaction between
  the scalar and electromagnetic fields:
\bearr                                                  \label{L}
     {\cal S} = \frac {1} {16\pi} \int \sqrt{-g} d^4 x
	\Big[ R + 2h(\phi) g^{\alpha\beta}\phi_{;\alpha}\phi_{;\beta}
\nnn \ \
	    - 2 V(\phi) - S(\phi) F_{\mu\nu}F^{\mu\nu} \Big],
\ear
  where $h > 0$ for a normal scalar field with positive kinetic energy
  and $ h < 0$ for a phantom one; the function $S(\phi) > 0$,
  characterizing the scalar-electromagnetic interaction, is arbitrary.
  The field equations are
\bear
    	4 h \nabla^\mu \nabla_\mu \phi
    		+ \eta S_\phi F\mn F \MN\eql 0,        	   \label{Ephi}
\\
        \nabla_\mu (S(\phi) F\MN) \eql 0,		   \label{EMax}
\\
       	R\mN - \half \delta\mN R \eql - T\mN,	  	   \label{EE}
\ear
  where $S_\phi \equiv  dS/d\phi$ and $T\mN$ is the SET:
\bearr                                                       
      T\mN = T\mN [\phi] + T\mN [F],
\nnn
      T\mN [\phi] =
      h(\phi)[ 2\phi_\mu \phi^\nu - \delta\mN \phi^{\alpha}\phi_{\alpha}]
		+ \delta\mN V(\phi),
\nnn
      T\mN [F] = S(\phi) \big[ -2F_{\mu\alpha}F^{\nu\alpha}
       	   + \half \delta\mN F_{\alpha\beta} F^{\alpha\beta}]. \label{SET}
\ear
  The general \ssph\ metric may be written in the form
\beq                                                         \label{ds1}
    ds^2 = \e^{2\gamma} dt^2 - \e^{2\alpha}du^2 - \e^{2\beta}d\Omega^2,
\eeq
  where $\gamma$, $\alpha$ and $\beta$ are functions of the radial
  coordinate $u$. There remains a coordinate freedom of choosing $u$.
  We also use the notation $r(u) \equiv \e^\beta$.

  We are going to study linear \sph\ perturbations of solutions to the field
  equations due to (\ref{L}). So, the metric has the form (\ref{ds1}) but
  now $\gamma(u,t) = \gamma(u) + \delta\gamma$ and similarly for other
  quantities, with small ``deltas''; for the scalar field we have $\phi
  (u,t) = \phi(u) + \delta\phi(u,t)$. The most general electromagnetic field
  compatible with spherical symmetry is described by the 4-potential
\beq
	A_\mu = \delta_\mu^0 A_0 + \delta_\mu^3 q_m \cos \theta + \d_\mu\Phi,
\eeq
  where $q_m$ is a magnetic charge and $\Phi$ an arbitrary function.
  The electromagnetic field equations give
\bear                                                    \label{F10}
	S(\phi) \e^{\alpha+2\beta+\gamma} F^{10} = q_e,
\ear
  where $q_e$ is an electric charge. We thus have
\beq
	F\mn F\MN = 2 \e^{-4\beta} (-q_e^2/S^2 + q_m^2).
\eeq
  For the SETs we obtain
\bear	  \nq						\label{SET-phi}
	T\mN [\phi] \eql \eps \e^{-2\alpha} \phi'^2
			\diag (1,\ -1,\ 1,\ 1) + \delta\mN V,
\\       \nq                                            \label{SET-F}
	T\mN [F] \eql \eta \e^{-4\beta} Q(\phi)
			\diag (1,\ 1,\ -1,\ -1).
\ear
  where $Q = Q(\phi) = q_e^2/S(\phi) + q_m^2 S(\phi)$.
  We will consider the equations governing linear radial perturbations of a
  \ssph\ solution to the field equations (\ref{Ephi})--(\ref{EE}),
  following the lines of \cite{we-book, sta1, sta2, sta3}.

  Preserving only linear terms with respect to time derivatives, we can
  write all nonzero components of the Ricci tensor as
\bear
     R^0_0 \eql                                         \label{R00}
     \e^{-2\gamma}(\ddot\alpha + 2\ddot\beta)
\nnn\ \ \
           -\e^{-2\alpha}[\gamma'' +\gamma'(\gamma'-\alpha'+2\beta')],
\yy
     R^1_1 \eql
     \e^{-2\gamma}\ddot\alpha                           \label{R11}
     - \e^{-2\alpha}[\gamma''+2\beta'' +\gamma'{}^2
\nnn \ \ \
     +2\beta'{}^2 - \alpha'(\gamma'+2\beta')],
\yy                                                      \label{R22}
     R^2_2 \eql R^3_3 = \e^{-2\beta}
          +\e^{-2\gamma}\ddot\beta
\nnn \ \ \
              -\e^{-2\alpha}[\beta''+\beta'(\gamma'-\alpha'+2\beta')],
\yy
     R_{01}\eql
        2[\dot\beta' + \dot{\beta}\beta'                        \label{R01}
                 -\dot{\alpha}\beta'-\dot{\beta}\gamma'],
\ear
  where dots and primes denote $\d/\d t$ and $\d/\d u$, respectively.

  The zero-order (static) scalar, ${0\choose 0}$, ${1\choose 1}$,
  ${2\choose 2}$ components of \eqs (\ref{EE}) are
\bearr      \nq                                            \label{e-phi0}
     2h[\phi'' + \phi'(\gamma'+2\beta'-\alpha')]
     		+ h'\phi' = \e^{2\alpha}P_\phi;
\yyy        \nq                                             \label{00-0}
     \gamma'' + \gamma'(\gamma'+2\beta'-\alpha') =
     				\e^{2\alpha}(-V + Q\e^{-4\beta});
\yyy        \nq                                                 \label{11-0}
     \gamma'' + 2\beta'' + \gamma'{}^2 + 2\beta'{}^2
            -\alpha'(\gamma'+2\beta')
\nnn \ \
	    =  -2h \phi'^2 +\e^{2\alpha}(-V + Q\e^{-4\beta});
\yyy        \nq                                                \label{22-0}
     -\e^{2\alpha-2\beta}
       + \beta'' + \beta'(\gamma'+2\beta'-\alpha')= -P\e^{2\alpha},
\ear
  where the subscript $\phi$ denotes $\d/\d\phi$ and
\beq                                               \label{def-P}
      P := V(\phi) + Q\e^{-4\beta},
\eeq
  The first-order perturbed equations (scalar, $R_{01}=\ldots$, and
  $R^2_2 = \ldots$) read
\bearr                                                     \label{e-phi1}
     2\e^{2\alpha-2\gamma} h \delta\ddot\phi - 2h[\df''
\nnn \
        +\df' (\gamma'+2\beta'-\alpha')
	+\phi'(\dg' + 2\db' -\da')]
\nnn \
	- 2\delta h[\phi'' + \phi'(2\beta'+\gamma'-\alpha')]
\nnn \
	- h' \df' - \phi'\delta h'
            +  \delta(\e^{2\alpha}P_\phi) =0,
\yyy                                                       \label{01-1}
    \delta\dot\beta' + \beta'\delta\dot\beta
        - \beta' \delta\dot\alpha - \gamma' \delta\dot{\beta}
                = - h \phi'\delta\dot\phi,
\yyy                                                       \label{22-1}
    \delta(\e^{2\alpha-2\beta})
            + \e^{2\alpha-2\gamma} \delta\ddot\beta
                    -\db'' -\db'(\gamma'+2\beta'-\alpha')
\nnn \qquad
        -\beta'(\dg' + 2\db' -\da')  = \delta(\e^{2\alpha} P),
\ear
  \eq (\ref{01-1}) may be integrated in $t$; since we are interested in
  time-dependent perturbations, we omit the appearing arbitrary
  function of $u$ describing static perturbations and obtain
\beq                                                       \label{01-1i}
    \db' + \db(\beta'-\gamma') - \beta'\da = -h \phi' \df.
\eeq

  Let us note that we have two independent forms of arbitrariness: one is the
  freedom of choosing a {\it radial coordinate\/} $u$, the other is a {\it
  perturbation gauge\/}, or, in other words, a reference frame in the
  perturbed space-time, which can be expressed in imposing a certain
  relation for $\da,\ \db$, etc. In what follows we will employ both kinds
  of freedom. All the above equations have been written in the most
  universal form, without coordinate or gauge fixing.

  Preserving the coordinate arbitrariness, we will now choose the simplest
  possible gauge $\db \equiv 0$. Then \eq (\ref{01-1i}) expresses $\da$ in
  terms of $\df$:
\beq                                  			      \label{da-df}
       \beta' \da = h(\phi) \phi' \df.
\eeq
  \eq (\ref{22-1}) expresses $\dg' - \da'$ in terms of $\da$ and $\df$:
\beq                                  \label{dg-df}
    \beta'(\dg'-\da') = 2 \e^{2\alpha-2\beta}\da
                        - \delta(\e^{2\alpha} P).
\eeq
  Substituting all this into (\ref{e-phi1}), we
  finally obtain the following wave equation:
\bearr                                                        \label{eq-df}
      \e^{2\alpha-2\gamma} \delta\ddot\phi
            -\df'' - \df' (\gamma'+ 2\beta'-\alpha'+ h'/h)
\nnn \inch
	    + U \df =0,
\yyy                                                          \label{def-U}
      U \equiv \Half \phi'{}^2
      		\biggl(\frac{h_\phi^2}{h^2}-\frac{h_{\phi\phi}}{h}\biggr)+ \e^{2\alpha}\biggl[
     	\frac{2h \phi'^2}{\beta'^2}(P - \e^{-2\beta})
\nnn \cm\cm
          + \frac{2\phi'}{\beta'} P_\phi + \frac{h_\phi}{2h^2} P_\phi
	         - \frac{P_{\phi\phi}}{2h}\biggr].
\ear
  This expression for $U$ directly generalizes the one obtained in
  \cite{sta3} for scalar-vacuum configurations. The latter is restored if we
  assume $h = \ep =\pm 1$ and $q_e = q_m =0$.

  Passing on to the ``tortoise'' coordinate $x$ introduced according to
\beq
       du/dx = \e^{\gamma-\alpha}                           \label{to_x}
\eeq
  and changing the unknown function $\df \mapsto \psi$ according to
\beq                                                  	     \label{to_psi}
       \df = \psi(x,t) \e^{-\eta}, \cm \eta' = \beta' + \frac{h'}{2h},
\eeq
  we reduce the wave equation to its canonical form, also called
  the master equation for radial perturbations:
\beq                                                        \label{wave}
       \ddot \psi - \psi_{xx} + \Veff (x)\psi =0,
\eeq
  (the index $x$ denotes $d/dx$), with the effective potential
\bearr \nq                                                    \label{Veff}
     \Veff (x) = \e^{2\gamma-2\alpha}
            [U + \eta''+ \eta'(\eta' + \gamma'-\alpha')].
\ear
  A further substitution
\beq                                                       \label{to_y}
       \psi (x, t) = y(x) \e^{i\omega t}, \cm \omega = \const,
\eeq
  which is possible because the background is static, leads to the
  \Schr-like equation
\beq                                                        \label{Schr}
      y_{xx} + [\omega^2 - \Veff(x)] y =0.
\eeq
  If there is a nontrivial solution to (\ref{Schr}) with $\im \omega <0$
  satisfying some physically reasonable conditions at the ends of the range
  of $u$ (in particular, the absence of ingoing waves), then the
  static system is unstable since $\df$ can exponentially grow with $t$.
  Otherwise our static system is stable in the linear approximation. Thus,
  as usual in such studies, the stability problem is reduced to a
  boundary-value problem for \eq (\ref{Schr}) --- see, e.g., \cite
  {we-book, sta1, sta2, sta3, we99, stepan04, stepan05, gon08, kb-kon, sha}.

  Note that all the above relations are written without fixing the background
  radial coordinate $u$.

  The gauge $\db = 0$ is technically the simplest one, but causes certain
  problems when applied to \whs\ and other configurations with throats. The
  reason is that the assumption $\db = 0$ leaves invariable the throat
  radius, while perturbation must in general admit its time dependence
  \cite{we99, gon08, we-book}. It may even seem that the emergence of a pole
  in $\Veff$ due to $\db$ in the denominator in (\ref{def-U}) is an artefact
  of the gauge. It turns out, however, that, by analogy with \cite{gon08,
  sta3, we-book}, \eq (\ref{Schr}) is in fact gauge-invariant, while $\df$
  is a representation of a gauge-invariant quantity in the gauge $\db =0$.

  Therefore, singularities of the effective potential $\Veff$ are of
  objective nature. The singularity at $\beta'=0$ (e.g., a throat) can be
  regularized \cite{gon08, sta3}, and moreover, it can be shown that
  regular solutions to the regularized equations describe regular
  perturbations of both the scalar field and the metric. It was this
  procedure that made it possible to prove the instability of anti-Fisher
  (Ellis type \cite{kb73, h_ell}) \whs\ \cite{gon08} and other scalar field
  configurations in \GR\ \cite{sta3, kb-kon}.

  The effective potential $\Veff$ also possesses singularities at the values
  of the radial coordinate where $h=0$, which exist in the cases where the
  function $h(\phi)$ in (\ref{L}) changes its sign. This happens in the
  framework of the trapped ghost concept. The existing experience
  \cite{sta1, sta2, gon08, sta3, kb-kon} shows that such singularities are
  in general an indication of instabilities, even if $\Veff \to \infty$ at
  such a singularity. As already mentioned, this was one of the reasons
  for our attempt to replace a ``trapped ghost'' in \wh\ and \bu\ models
  with an ``invisible'' one.

\end{document}